\documentclass[12pt]{article}

\usepackage{epsfig}
\usepackage{amssymb,amsmath}

\newcommand{\bea}{\begin{eqnarray}}
\newcommand{\eea}{\end{eqnarray}}

 \makeatletter
\def\alt{\mathrel{\mathpalette\gl@align<}}
\def\agt{\mathrel{\mathpalette\gl@align>}}
\def\gl@align#1#2{\lower.6ex\vbox{\baselineskip\z@skip\lineskip\z@
\ialign{$\m@th#1\hfil##\hfil$\crcr#2\crcr\sim\crcr}}} \makeatother

\begin{document}
\begin{flushright}
BA-07-31
\end{flushright}
\vspace*{1.0cm}

\begin{center}
\baselineskip 20pt {\Large\bf Window For Higgs Boson Mass \\
From Gauge-Higgs Unification
 } \vspace{1cm}

{\large Ilia Gogoladze$^{a,}$\footnote{ On a leave of absence from:
Andronikashvili Institute of Physics, GAS, 380077 Tbilisi, Georgia.
\\ \hspace*{0.5cm} E-mail:
ilia@physics.udel.edu}, Nobuchika Okada$^{b,c,}$\footnote{ E-mail:
okadan@post.kek.jp} and Qaisar Shafi$^{a,}$\footnote{ E-mail:
shafi@bartol.udel.edu} } \vspace{.5cm}

{\baselineskip 20pt \it
$^a$Bartol Research Institute, Department of Physics and Astronomy, \\
University of Delaware, Newark, DE 19716, USA \\
\vspace{2mm} $^b$Department of Physics, University of Maryland,
College Park,  MD 20742, USA \\
\vspace{2mm} $^c$Theory Division, KEK, Tsukuba 305-0801, Japan }
\vspace{.5cm}

\vspace{1.5cm} {\bf Abstract}
\end{center}

We consider six dimensional gauge models compactified on the orbifold
 $T^2/Z_N$ ($N=2,3,4,6$) such that the Standard Model (SM) Higgs
 doublet arises from the extra-dimensional components of the gauge field.
For $\Lambda \leq 10^{19}$ GeV, where $\Lambda$ denotes the
compactification scale,
 we obtain $114.4$ GeV $\leq m_H \leq 164$ GeV for the SM Higgs boson mass.
We also consider gauge-Higgs-top and gauge-Higgs-bottom Yukawa
 unification which respectively yield
 $m_H = 131^{+4}_{-5}$ GeV and $m_H = 150^{+2}_{-2}$ GeV
 for a top quark pole mass $M_t =170.9^{+1.8}_{-1.8}$ GeV.
As a special case we recover the result $m_H \leq 132$ GeV
 previously obtained for five dimensional models.

\thispagestyle{empty}

\newpage

\addtocounter{page}{-1}

\baselineskip 18pt

In a recent paper \cite{GOS}, hereafter called I,
 we investigated five dimensional (5D) gauge models
 compactified on the orbifold $S^1/Z_2$ in which
 the zero mode of the fifth component of the gauge field
 could be identified with the SM Higgs doublet $H$.
The five dimensional gauge invariance requires that
 the Higgs quartic coupling vanish at tree level.
By imposing this condition at the compactification scale
 $\Lambda$ \cite{GHUcond}, the SM Higgs boson mass was estimated
 using two-loop renormalization group equations (RGE).
For $10^6$ GeV$\leq \Lambda \leq 10^{19}$ GeV,
 and for a top quark pole mass of $170.9 \pm 1.8$ GeV \cite{Tevatron},
 the mass is in the range 114.4 GeV $\leq m_H \leq 132$ GeV.
[In I the upper bound was found to be 129 GeV. A more careful
treatment here of the top
 quark pole mass yields the slightly larger value of 132 GeV.]
In the SM with the standard particle content, the SU(2) gauge coupling
 and the top quark Yukawa coupling have the same magnitude
 at scales of order $10^8$ GeV.
If the latter scale is identified with $\Lambda$,
 one obtains $m_H = 117 \pm 4$ GeV \cite{GOS}.
It is amusing to note that the Higgs boson mass predictions
 in this 5D model have a great deal of overlap with
 the minimal supersymmetric Standard Model (MSSM) prediction
 for the mass of the lightest CP-even Higgs boson.
The uncertainty in the Higgs mass predictions are largely
 due to the experimental uncertainty in the determination
 of the top quark mass.

In this paper we extend our earlier results in I
 by considering 6D gauge-Higgs unification (GHU) models
 compactified on the orbifold $T^2/Z_N$, with $N=2,3,4$ and 6.
For $N=2$, two SU(2) doublets appear under appropriate boundary
 conditions as the zero modes of the extra-dimensional components
 of the gauge field \cite{6DGHU1}.
Since our goal here is to predict the mass of the SM Higgs boson $H$,
 we will assume that a suitable linear combination of these two
 doublets corresponds to $H$, while the orthogonal combination
 acquires mass of order $\Lambda$.
The six dimensional gauge invariance determines the quartic tree
 level coupling of $H$ in terms of the SU(2) gauge coupling
 and $\tan \beta$, where $\tan \beta$ is defined, as usual,
 as the ratio of the vacuum expectation value of
 the two Higgs doublets.
With this value of the quartic coupling as the boundary condition
 at a given scale $\Lambda$, and for a given $\tan \beta$,
 the Higgs boson mass is estimated using two-loop RGEs.
We find 114.4 GeV$ \leq m_H \leq 164$ GeV,
 for $\Lambda \leq 10^{19}$ GeV,
 with $0 \leq \tan \beta < \infty$.
Compactification on  $T^2/Z_N$ orbifolds with $N=3,4,6$ leaves only
 a single Higgs doublet \cite{6DGHU2}, as desired.
The Higgs mass for these cases is realized in the limit
 $\tan \beta =0$ or $\infty$.
We also discuss gauge-Higgs-Yukawa unification for which a more precise
 prediction for $m_H$ is found.
Thus, $m_H = 131$ GeV (150 GeV) with top (bottom) quark unification
 for a top quark pole mass $M_t=170.9$ GeV.
Finally, for $\tan \beta = 1$, we recover our earlier result for
 $m_H$($\leq$132 GeV) obtained in I with 5D gauge-Higgs unification.

We begin with a very brief review of the basic structure of 6D GHU models.
As a simple example, consider a SU(3) GHU model
 compactified on the orbifold $T^2/Z_N$ ($N=2,3,4,6$)
 \cite{6DGHU1, 6DGHU2}.
Note that at least one additional U(1) factor is needed
 to recover the 4D electroweak gauge symmetry.
However, this Abelian factor will not affect our discussion below.
The six dimensional Lagrangian for the SU(3) gauge field,
 $ A_{\hat{\mu}} = A^a_{\hat{\mu}} \lambda^a/2$,
 with $\lambda^a$ the Gell-Mann matrix, is given by
\bea
  {\cal L} =
 -\frac{1}{2} {\rm tr} \left[
  F^{\hat{\mu} \hat{\nu}} F_{\hat{\mu} \hat{\nu}}
  \right],
\label{6DL}
\eea
where $ F_{\hat{\mu} \hat{\nu}}
  =  \partial_{\hat{\mu}} A_{\hat{\nu}}
   - \partial_{\hat{\nu}} A_{\hat{\mu}}
   -i g_6 \left[ A_{\hat{\mu}}, A_{\hat{\nu}} \right] $,
  $g_6$ is the 6D gauge coupling, and
  $\hat{\mu}, \hat{\nu}=0,1,2,3,5,6$.
In the following, we use the notation
 $\mu, \nu=0,1,2,3$ and $M, N =5,6$
 for the usual four dimensions and the extra two dimensions, respectively.
In terms of four dimensional effective theory,
 $A_\mu$ corresponds to the vector field
 while $A_M$ denotes the scalar fields.

In order to describe the model,
 it is useful to introduce a complex coordinate
 $ z = (x^5 + i x^6)/\sqrt{2} $  on the torus $T_2$ and
 a corresponding gauge field $A_z = (A_5 - i A_6)/\sqrt{2}$.
Under the $T^2/Z_N$ orbifold transformation, $z \rightarrow \tau z$
 with $\tau=e^{i 2 \pi/N}$,
 the transformation law, which keeps the action invariant,
 for the four dimensional and extra-dimensional components
 of the gauge field is given as
\bea
 A_\mu(x^\mu, \tau z) &=&
    \hat{P} A_\mu(x^\mu, x^5, x^6) \hat{P}^\dagger  ,
\nonumber \\
 A_z (x^\mu, \tau z)   &=&
   \tau^{-1} \hat{P} A_z (x^\mu, z) \hat{P}^\dagger ,
\label{trans}
\eea
with $\hat{P}={\rm diag}(\tau, \tau, 1)$.
With this boundary condition, the SU(3) gauge symmetry
 is broken down to SU(2)$\times$U(1).

The zero modes of the vector (gauge) fields can be
 explicitly written as
\bea
 A_{\mu}&=&\frac{1}{2 \sqrt{V_2}}
 \left(\begin{array}{ccc}
  A_{\mu}^{(3)}+\frac{1}{\sqrt{3}}A_{\mu}^{(8)} & A_{\mu}^{(1)}-i A_{\mu}^{(2)} & 0\\
  A_{\mu}^{(1)}+i A_{\mu}^{(2)} & -A_{\mu}^{(3)}+\frac{1}{\sqrt{3}}A_{\mu}^{(8)} & 0 \\
  0 & 0 & -\frac{2}{\sqrt{3}}A_{\mu}^{(8)} \end{array}\right)
\nonumber \\
\nonumber & \mbox{ } & \\
&=& \frac{1}{2 \sqrt{V_2}}\left(\begin{array}{ccc}
  W_{\mu}^{(3)}+\frac{1}{\sqrt{3}}B_{\mu}^{(8)} & \sqrt{2}W_{\mu}^+ & 0\\
  \sqrt{2}W_{\mu}^- & -W_{\mu}^{(3)}+\frac{1}{\sqrt{3}}B_{\mu}^{(8)} & 0 \\
  0 & 0 & -\frac{2}{\sqrt{3}}B_{\mu}^{(8)} \end{array}\right) ,
\label{zero-gauge}
\eea
 where $V_2$ is the volume of the extra dimensions
 and we have used the usual SU(2) notation.
The off-diagonal components of $A_z$, which form the SU(2) doublet fields,
 include candidates for zero-modes,
\bea
  A_z = \frac{1}{2 \sqrt{V_2}}
  \left( \begin{array}{cc}
    0 &  \sqrt{2} H_2 \\
    \sqrt{2} H_1^T &  0
 \end{array}\right),
\label{zero-scalar}
\eea
 where we have defined the two independent doublet fields as
\bea
 H_1 =
     \frac{1}{\sqrt{2}} \left( \begin{array}{c}
       A_z^{(4)} + i A_z^{(5)} \\
       A_z^{(6)} + i A_z^{(7)}
     \end{array}\right), \;
 H_2 =
     \frac{1}{\sqrt{2}} \left( \begin{array}{c}
      A_z^{(4)} - i A_z^{(5)} \\
      A_z^{(6)} - i A_z^{(7)}
     \end{array}\right) .
\label{defHiggs1}
\eea
According to Eq.~(\ref{trans}), the transformation law
 for these SU(2) doublets is explicitly written as
\bea
 H_1(x^\mu, \tau z) &=& \tau^{-2} H_1(x^\mu, z) ,
\nonumber \\
 H_2(x^\mu, \tau z) &=&  H_2(x^\mu, z) .
\eea
Therefore, under this twisted boundary condition,
 the doublet $H_2$ always has a zero-mode,
 while the doublet $H_1$ has a zero-mode
 only when $N=2$ and $ \tau^{-2}=1$ .

Substituting these various expressions in Eq.~(\ref{6DL}) and
integrating over
 the extra-dimensional coordinates $(x_5, x_6)$,
 we obtain the Lagrangian of SU(2)$\times$U(1)
 with two Higgs doublets in four dimensions.
In particular, the Higgs potential is obtained from the term,
\bea
V &=& \int d x^5 d x^6 \frac{1}{2} {\rm tr}
   \left[ F^{MN} F_{MN} \right]
  = \int d x^5 d x^6 \; {\rm tr}
   \left[ (F_{56})^2 \right]  \nonumber \\
 &=&  - g_6^2 \int d x^5 d x^6 \; {\rm tr} \left[ A_5, A_6 \right ]^2
=  g_6^2 \int d x^5 d x^6 \;
   {\rm tr} \left[ A_z, A_z^\dagger \right]^2,
\label{HPot}
\eea
where we have used
 $F_{MN} = \partial_M A_N - \partial_N A_M -i g_6 [A_M,A_N]
 = -i g_6 [A_M,A_N] $ for the zero modes,
 and the definition $A_z=(A_5 - i A_6)/\sqrt{2}$
 in the last expression.

In the $N=2$ case, the model provides two scalar SU(2) doublets
 as zero modes.
We define the up-type and down-type Higgs doublets as
\bea
  H_u = H_2^0 , \; \;  H_d = - i \sigma_2 H_1^0 ,
\label{defHiggs2}
\eea
where $H_{1,2}^0$ denotes the zero-mode of $H_{1,2}$.
Substituting the explicit matrix expression
 of Eq.~(\ref{zero-scalar}) together with Eqs.~(\ref{defHiggs1})
 and (\ref{defHiggs2}) into Eq.~(\ref{HPot}),
 we obtain the following Higgs potential \cite{6DGHU1}
\bea
V = \frac{g^2}{2} \left( |H_d|^2 - |H_u|^2 \right)^2
   + \frac{g^2}{2}
     (H_u^\dagger H_d) (H_d^\dagger H_u) ,
\label{quartic} \eea
where $g = g_6/\sqrt{V_2}$ is the SU(2) gauge
coupling. The quadratic mass term is forbidden at tree level
 by the six dimensional gauge invariance.
This formula corresponds to the D-term potential
 in the MSSM with the identification $g^{\prime 2} = 3 g^2$
 for the U(1)$_Y$ gauge coupling.

In the case of $N=3,4,6$,
 only a single Higgs doublet field emerges as the zero-mode of $H_2$,
 which is identified as the SM Higgs doublet.
The quartic Higgs potential in this case is given by \cite{6DGHU2},
\bea
  V= \frac{1}{2} g^2 |H|^4 ,
\eea
where we have defined the SM higgs doublet as $H=H_2^0$.

As has been explicitly shown in Ref.~\cite{GHUcond} (for a 5D GHU model),
 the effective SM Higgs quartic coupling calculated
 in a given GHU model coincides with the one
 generated through the RGEs in the SM by imposing
 a special boundary condition at the compactification scale.
This boundary condition is the gauge-Higgs condition,
 where the quartic coupling at the compactification scale is
 set to be the tree level one required
 by the higher dimensional gauge invariance.
This treatment is natural from an effective field theoretical point of view,
 where Kaluza-Klein modes should decouple at low energy,
 and (dimensionless) couplings in the low energy effective theory
 should be matched with the one from high energy theory
 at the compactification scale.
Corrections to the Higgs mass squared in GHU models, on the other hand,
 are very much dependent on the particle contents and
 imposed boundary conditions (see, for instance \cite{GHUcond}).
In this paper we will therefore treat the Higgs mass squared
 as a free parameter in the low energy effective theory, to be
 suitably adjusted to yield the desired electroweak symmetry breaking.
Once the electroweak symmetry breaking is correctly achieved,
 the Higgs boson mass is determined by its quartic coupling
 at the electroweak scale.

We are now ready to discuss the physical Higgs boson mass $m_H$. For
compactification on  $T^2/Z_2 $, only one combination of
 the Higgs doublets is assumed to be light and identified
 as the SM Higgs doublet, while the orthogonal combination has
 a compactification mass scale and decouples at low energies.
Under this assumption, the  Higgs doublets $H_u, H_d$ are described
 in terms of the SM(-like) Higgs doublet ($H$),
 the heavy Higgs doublet ($\tilde{H}$) and
 $\tan\beta= \langle H_u \rangle/\langle H_d  \rangle$,
\bea
 H_u &=& H \sin \beta  + \tilde{H} \cos \beta,
\nonumber  \\
i \sigma_2 H_d^* &=& H \cos \beta  - \tilde{H} \sin \beta. \eea

The quartic coupling of the SM Higgs boson is read off
 from Eq.~(\ref{quartic}) as
\bea
  V = \frac{1}{2} g^2 \cos^2(2 \beta) |H|^4 .
\label{HiggsQuartic}
\eea
Note that the one Higgs doublet case corresponding to
 compactification on the orbifold $T^2/Z_N$ with $N \neq2$,
 can be realized as the limit $\tan \beta=0/\infty$,
 or equivalently $\cos^2(2 \beta)=1$.
In this case $m_H=2m_W$ at tree level \cite{6DGHU2}.
Also, for $\tan \beta=1$ or $\cos(2 \beta)=0$, we recover
 the condition analyzed in Ref.~\cite{GOS} for 5D GHU models.
Therefore, the 6D GHU model with varying $\tan\beta$
 provides the general results for the SM Higgs boson mass
 in this class of models.

Imposing the gauge-Higgs condition for the Higgs quartic coupling,
 $\lambda = g^2 \cos^2(2 \beta)$, at a given compactification scale
 $\Lambda$, and for a given $\tan \beta$, we solve
 the two loop RGEs \cite{RGE}, to obtain the Higgs boson mass.
Namely,
\bea
 m_H (m_H) = \sqrt{\lambda(m_H)} \; v .
\eea

For the  three SM gauge couplings, we have
\bea
 \frac{d g_i}{d \ln \mu} =  \frac{b_i}{16 \pi^2} g_i^3 +\frac{g_i^3}{(16\pi^2)^2}
\sum_{j=1}^3B_{ij}g_j^2, \label{gauge} \eea
 where $ \mu$ is the renormalization scale,
$g_i$ ($i=1,2,3$) are the SM  gauge couplings  and
\begin{equation}
b_i = \left(\frac{41}{10},-\frac{19}{6},-7\right),~~~~~~~~
 { b_{ij}} =
 \left(
  \begin{array}{ccc}
  \frac{199}{50}& \frac{27}{10}&\frac{44}{5}\\
 \frac{9}{10} & \frac{35}{6}&12 \\
 \frac{11}{10}&\frac{9}{2}&-26
  \end{array}
 \right).
\end{equation}
The top quark pole mass is taken to be
 $M_t= 170.9 \pm 1.8$ GeV,  \cite{Tevatron},
with
 $(\alpha_1, \alpha_2, \alpha_3)=(0.01681, 0.03354, 0.1176)$
 at the Z-pole ($M_Z$) \cite{PDG}.
For the top Yukawa coupling $y_t$, we have \cite{RGE},
\bea \label{ty}
 \frac{d y_t}{d \ln \mu}
 = y_t  \left(
 \frac{1}{16 \pi^2} \beta_t^{(1)} + \frac{1}{(16 \pi^2)^2} \beta_t^{(2)}
 \right).
\eea
Here the one-loop contribution is
\bea
 \beta_t^{(1)} =  \frac{9}{2} y_t^2 -
  \left(
    \frac{17}{20} g_1^2 + \frac{9}{4} g_2^2 + 8 g_3^2
  \right),
\label{topYukawa-1}
\eea
while the two-loop contribution is given by
\bea
\beta_t^{(2)} &=&
 -12 y_t^4 +   \left(
    \frac{393}{80} g_1^2 + \frac{225}{16} g_2^2  + 36 g_3^2
   \right)  y_t^2  \nonumber \\
 &+& \frac{1187}{600} g_1^4 - \frac{9}{20} g_1^2 g_2^2 +
  \frac{19}{15} g_1^2 g_3^2
  - \frac{23}{4}  g_2^4  + 9  g_2^2 g_3^2  - 108 g_3^4 \nonumber \\
 &+& \frac{3}{2} \lambda^2 - 6 \lambda y_t^2 .
\label{topYukawa-2}
\eea
In solving Eq.~(\ref{ty})
 from $M_t$ to the compactification scale $\Lambda$,
 the initial top Yukawa coupling at $\mu=M_t$
 is determined from the relation
 between the pole mass and the running Yukawa coupling
 \cite{Pole-MSbar}, \cite{Pole-MSbar2},
\bea
  M_t \simeq m_t(M_t)
 \left( 1 + \frac{4}{3} \frac{\alpha_3(M_t)}{\pi}
          + 11  \left( \frac{\alpha_3(M_t)}{\pi} \right)^2
          - \left( \frac{m_t(M_t)}{2 \pi v}  \right)^2
 \right),
\eea
 with $ y_t(M_t) = \sqrt{2} m_t(M_t)/v$, where $v=246.2$ GeV.
Here, the second and third terms in the parenthesis correspond to
 one- and two-loop QCD corrections, respectively,
 while the fourth term comes from the electroweak corrections at one-loop level.
The numerical values of the third and fourth terms
 are comparable (signs are opposite).
The electroweak corrections at two-loop level and
 the three-loop QCD corrections \cite{Pole-MSbar2},
 are of comparable and sufficiently small magnitude \cite{Pole-MSbar2}
 to be safely ignored.

The RGE  for the Higgs quartic coupling is given by \cite{RGE},
\bea
\frac{d \lambda}{d \ln \mu}
 =   \frac{1}{16 \pi^2} \beta_\lambda^{(1)}
   + \frac{1}{(16 \pi^2)^2}  \beta_\lambda^{(2)},
\label{self}
\eea
with
 \bea
 \beta_\lambda^{(1)}=
 12 \lambda^2 -
 \left(  \frac{9}{5} g_1^2+9 g_2^2  \right) \lambda
 + \frac{9}{4}
 \left(
 \frac{3}{25} g_1^4 + \frac{2}{5} g_1^2 g_2^2 +g_2^4
 \right)
 + 12 y_t^2 \lambda -12 y_t^4 ,
\label{self-1}
\eea
and
\bea
  \beta_\lambda^{(2)} &=&
 -78 \lambda^3  + 18 \left( \frac{3}{5} g_1^2 + 3 g_2^2 \right) \lambda^2
 - \left( \frac{73}{8} g_2^4  - \frac{117}{20} g_1^2 g_2^2
 + \frac{2661}{100} g_1^4  \right) \lambda - 3 \lambda y_t^4
 \nonumber \\
 &+& \frac{305}{8} g_2^6 - \frac{289}{40} g_1^2 g_2^4
 - \frac{1677}{200} g_1^4 g_2^2 - \frac{3411}{1000} g_1^6
 - 64 g_3^2 y_t^4 - \frac{16}{5} g_1^2 y_t^4
 - \frac{9}{2} g_2^4 y_t^2
 \nonumber \\
 &+& 10 \lambda \left(
  \frac{17}{20} g_1^2 + \frac{9}{4} g_2^2 + 8 g_3^2 \right) y_t^2
 -\frac{3}{5} g_1^2 \left(\frac{57}{10} g_1^2 - 21 g_2^2 \right)
  y_t^2  - 72 \lambda^2 y_t^2  + 60 y_t^6.
\label{self-2}
\eea

In Figure~1, we plot the Higgs boson mass $m_H$
 as a function of the compactification scale
 for a given $\tan \beta$ with an input top quark pole mass
 $M_t= 170.9 \pm 1.8$ GeV.
Each set of three lines (in red, black and blue) corresponds
 to $M_t=172,7$, $170.9$ and $169.1$ GeV, from top to bottom.
The upper three lines (in red, black and blue) are the results
 for the gauge-Higgs condition $\lambda=g^2$
 (with $\tan\beta=0/\infty$ or equivalently $|\cos(2\beta)|=1$),
 which also correspond to the results
 in 6D GHU models with the orbifold compactifications $T^2/Z_N$
 ($N \neq 2$), as previously mentioned.
The lower three lines are the results for the gauge-Higgs condition
 $\lambda= 0$ (with $\tan \beta =1$ or equivalently $\cos(2\beta)=0$),
 which are exactly the same ones presented in Ref.~\cite{GOS}.
For general values of $\tan \beta$ (or $\cos(2\beta)$),
 the resultant Higgs mass should appear between the top and bottom lines.
Therefore, in this class of models, the Higgs mass is predicted in the range
 114.4 GeV$\leq m_H \leq$ 164 GeV,
 for a cutoff scale lower than the Planck scale.
The range is narrow compared to the one obtained
 from the stability and triviality bounds on
 the Higgs quartic coupling \cite{stability1}.

In the GHU model with fermions in the bulk one would expect
unification
 of gauge and Yukawa interactions at the compactification scale
 $\Lambda$.
This is clearly not plausible for fermions in the first two generations.
(To realize the hierarchy of fermion masses of the SM,
 a more elaborate GHU model must  be considered.
 There have been various efforts along this direction \cite{Csaki}.)
However, for a fermion in the third generation,
 its running Yukawa coupling and the running SU(2) gauge coupling
 can meet at some scale, which depends on $\tan \beta$,
 and it would be natural to identify this 'merger' point
 with the compactification scale. Let us first assume that
 this holds for the top quark Yukawa coupling.
 We have \bea
  {\cal L}_Y =
 - g \; \overline{q_{3L}} H_u t_R ,
\eea
where $q_{3L}$ is the quark doublet of the third generation,
 and the up-type Higgs $H_u = H \sin \beta + \tilde{H}\cos \beta$.
The SM top Yukawa coupling is defined as $y_t = g \sin\beta$. We can
fix the compactification scale as the point
 where the relation, $y_t(\Lambda) = g (\Lambda) \sin\beta$,
 is satisfied, with a given $\tan\beta$ (see Figure~2).
The compactification scale as a function of $\tan \beta$
 is depicted in Figure~3 for input top quark pole masses,
 $M_t=$169.1, 170.9 and 172.8 GeV.
Figure~4 shows the Higgs mass as a function of
 the compactification scale which has been fixed
 for a given $\tan\beta$.
This observation allows us to realize gauge-Higgs and gauge-top Yukawa
 coupling unification at $\Lambda$, \cite{GHY},
 and to narrow down the Higgs boson mass window from Figure~1
 to $125$ GeV$\leq m_H \leq 158$ GeV in Figure~4.

In $T^2/Z_2$ models with large $\tan \beta$,
 the bottom quark and tau Yukawa couplings can be large,
 and it is possible for the running bottom quark (or tau)
 Yukawa coupling to meet the running gauge coupling at a high scale.
As an example, we examine the unification of gauge and bottom quark
 Yukawa couplings, with
\bea
  {\cal L}_Y =
    - g \; \overline{q_{3L}} H_d b_R .
\eea with the down-type Higgs defined by Eq. (11),
 so that the SM bottom Yukawa coupling is given as $y_b = g \cos\beta$.

For our analysis, we employ the two-loop RGE of bottom Yukawa coupling \cite{RGE},
\bea \label{by}
 \frac{d y_b}{d \ln \mu}
 = y_b  \left(
 \frac{1}{16 \pi^2} \beta_b^{(1)} + \frac{1}{(16 \pi^2)^2} \beta_b^{(2)}
 \right).
\eea
Here the one-loop contribution is
\bea \label{bYukawa-1}
 \beta_b^{(1)} =
 \frac{3}{2} y_t^2 -
   \left( \frac{1}{4} g_1^2 + \frac{9}{4} g_2^2 + 8 g_3^2
 \right),
\eea
while the two-loop contribution is given by
\bea  \label{bYukawa-2}
\beta_b^{(2)} &=&
  -\frac{1}{4}  y_t^4 +  \left(
 \frac{91}{80}  g_1^2  + \frac{99}{16} g_2^2  + 4 g_3^2 \right) y_t^2
\nonumber \\
&-& \frac{127}{600} g_1^4  - \frac{23}{4}  g_2^4  - 108 g_3^4
  - \frac{27}{20}  g_1^2 g_2^2 + \frac{31}{15} g_1^2 g_3^2
  + 9  g_2^2 g_3^2 .
\eea
In this RGE, we have considered only terms involving
 gauge couplings and top Yukawa coupling
 in $\beta_b^{(1)}$ and $\beta_b^{(2)}$, as a good approximation.
Solving this RGE with the running bottom quark mass
 at the Z-pole as $m_b(M_Z)=3$ GeV \cite{FuKo}, for simplicity,
 we can fix the compactification scale as the point
 where the relation, $y_b(\Lambda) = g (\Lambda) \cos \beta$,
 is satisfied, with a given $\tan\beta$ (see Figure~5).
The compactification scale as a function of $\tan \beta$
 is depicted in Figure~6
 for input top quark pole masses, $M_t=$169.1, 170.9 and 172.8 GeV.
Figure~7 shows the Higgs mass as a function of
 the compactification scale which has been fixed
 for a given $\tan\beta$.
The resultant Higgs boson mass is found in
  the range $146$ GeV$\leq m_H \leq 164$ GeV.
We have verified that
 our results for the Higgs boson mass show only a very mild
 dependence on the few percent uncertainty present
 in the bottom quark mass.

Finally, it is tempting to simultaneously impose both gauge
 coupling and gauge-Yukawa unification at the compactification scale.
Although the three SM gauge couplings do not meet with
 a canonical normalization of $5/3$  for U(1)$_Y$,
 a different choice, for example, $4/3$, can lead
 to gauge coupling unification
 at $M_\text{GUT} = 4 \times 10^{16}$ GeV \cite{NCY1, NCU}.
If we identify this gauge coupling unification scale
 with the compactification scale $\Lambda$,
 the SM Higgs boson mass and $\tan\beta$ can both be fixed.
For a top quark pole mass $M_t=170.9^{+1.8}_{-1.8}$ GeV,
 we find $m_H=131^{+4}_{-5}$ GeV and $\tan \beta = 1.32^{+0.09}_{-0.08}$
 in the case of gauge-top Yukawa coupling unification,
 while $m_H = 150^{+2}_{-2}$ GeV and $\tan \beta = 88.2^{-0.3}_{+0.4}$
 for gauge-bottom Yukawa coupling unification.

In conclusion, we have considered gauge-Higgs unification models
 in six dimensions with  $T^2/Z_N$ orbifold compactification,
 such that up to two SM Higgs doublets emerge as the extra-dimensional
 components of the higher dimensional gauge field.
The effective quartic Higgs coupling at tree level
 is determined from higher dimensional gauge invariance,
 that should be matched with the one from the low energy effective
 theory below the compactification scale (the gauge-Higgs condition).
In the $N=2$ case, the model provides two scalar doublets
 as zero-modes of the extra-dimensional components of
 the higher dimensional gauge field.
We have assumed that one linear combination (the SM Higgs doublet)
survives below the compactification scale,
 and that the low energy effective theory coincides with the Standard Model.
Imposing the gauge-Higgs condition and solving the RGEs in the SM,
 we have obtained the Higgs boson mass
 in the range, 114.4 GeV$\leq m_H \leq$164 GeV,
 for given compactification scale and $\tan\beta$,
 with the top quark pole mass $M_t=170.9 \pm 1.8$ GeV.
The Higgs boson mass in 5D GHU models
 with $S^1/Z_2$ orbifold compactification and for
 6D GHU models compactified on $T^2/Z_N$,
 with $N \neq 2$, can be obtained by setting
 $\tan \beta=1$ and $\tan \beta =0/\infty$, respectively.
Interestingly, these two limits fix the lower and upper boundaries
 of the resultant Higgs boson mass window.
Therefore, the Higgs boson mass we have found in this letter
 is the general result in this class of models.
Imposing further conditions such as gauge-Yukawa coupling unification
 enables us to confine the Higgs boson mass in a much more narrow range.

Finally, we offer some comments concerning non-baryonic dark matter
whose presence has been established from various observations of the
present universe. Except for the Higgs sector,
 the gauge-Higgs unification model shares the same structure
 as the Universal Extra Dimension (UED) model \cite{Antoniadis:1990ew, UED},
 in which, due to a conserved KK parity, the lightest KK mode
 is a plausible dark matter candidate \cite{UED-DM}.
If the compactification scale is around 1 TeV,
 KK dark matter as a thermal relic is found to be
 consistent with the observed dark matter density
 in the present universe \cite{LKP}.
In our case, for $|\cos (2 \beta)| \geq 0.71$
 (see Eq.~(\ref{HiggsQuartic})),
 the compactification scale can be somewhat less than 1 TeV
 with the resultant Higgs boson mass $m_H \geq 114.4$ GeV,
 so that the KK dark matter scenario can be realized.
If the compactification scale exceeds the unitarity limit
 on the mass of cold dark matter as a thermal relic,
 a superheavy KK dark matter scenario may still be realized
 through another mechanism, such as its production
 through inflaton decay.

\section*{Acknowledgments}
We  thank Chin-Aik (Jason) Lee for very helpful discussions. N.O.
would like to thank the Particle Theory Group
 of the University of Delaware for hospitality during his visit.
This work is supported in part by
 the DOE Grant \# DE-FG02-91ER40626 (I.G. and Q.S.),
 and
 the Grant-in-Aid for Scientific Research from the Ministry
 of Education, Science and Culture of Japan,
 \#18740170 (N.O.).


\vspace*{4.0cm}

%
\begin{figure}[t, width=12cm, height=8cm]
\begin{center}
{\includegraphics[height=7cm]{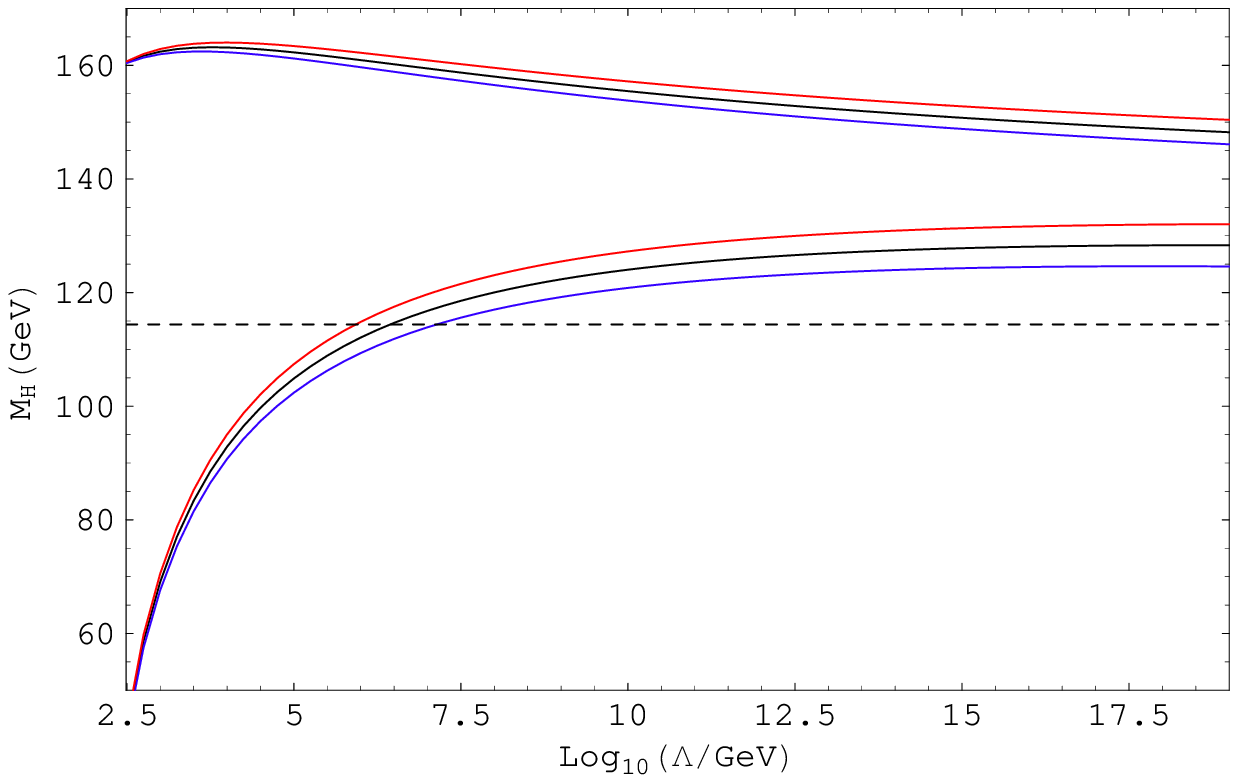}}
\end{center}
\caption{\small
  Higgs boson mass prediction versus
 the compactification scale for a given $\cos(2 \beta)$.
The upper three lines (in blue, black and red) correspond,
 from bottom to top, to input top quark pole masses,
 $M_t=$169.1, 170.9 and 172.7 GeV,
 for $\tan \beta =0/\infty$ or equivalently $\cos^2(2 \beta)=1$.
The lower three lines (in blue, black and red) correspond,
 from bottom to top, to input top quark pole masses,
 $M_t=$169.1, 170.9 and 172.7 GeV,
 for $\tan \beta=1$ or equivalently $\cos^2(2 \beta)=0$.
The lower lines show the same results
 as in 5D GHU models \cite{GOS}.
The horizontal line shows the current Higgs boson mass bound,
 $m_H \geq 114.4$ GeV, from LEP2 \cite{LEP2}.
}
%
\end{figure}

\begin{figure}[h, width=12cm, height=8cm]
\begin{center}
{\includegraphics[height=7cm]{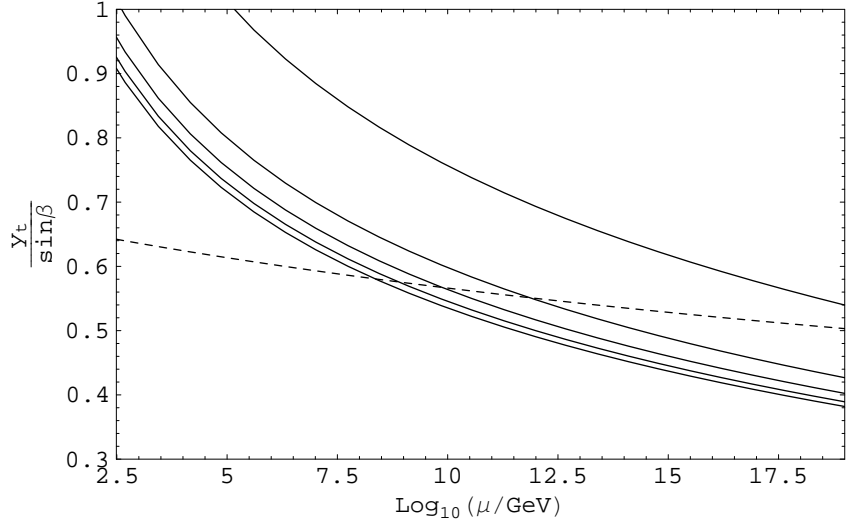} }
\end{center}
\caption{ \small
Plot of the running SU(2) gauge coupling $g(\mu)$ (dashed line)
 and the running top Yukawa coupling (divided by a given $\sin\beta$),
 $y_t(\mu)/\sin \beta$ (solid line), versus
 $\mbox{Log}_{10}( \mu /\mbox{GeV})$ for a given $\tan\beta$.
Solid lines correspond to $\tan \beta=1$, 2, 3, 5, and 30, respectively,
 from top to bottom.
Here we took top quark pole mass $M_t=170.9$ GeV.
}
%
\end{figure}

\begin{figure}[bt,width=12cm, height=8cm]
\begin{center}
{\includegraphics[height=7cm]{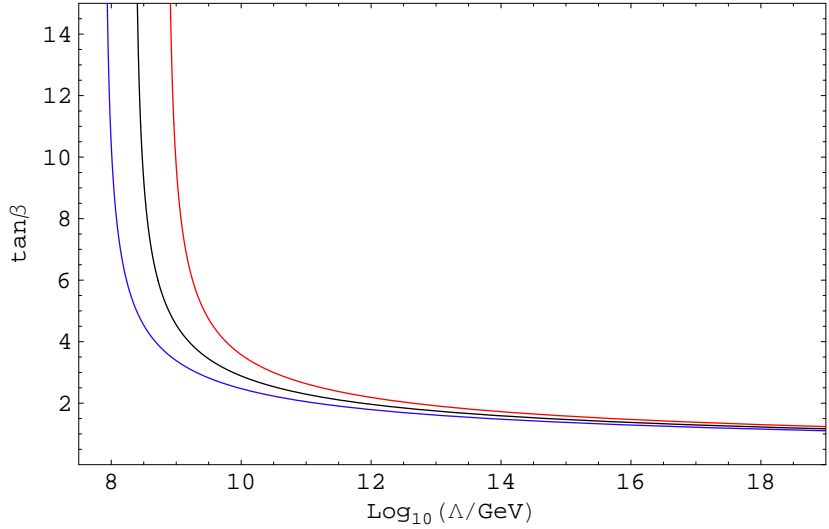} }
\end{center}
\caption{ \small
Plot of $\tan \beta$ versus the compactification scale,
 $\mbox{Log}_{10}( \Lambda /\mbox{GeV})$,
 determined as the merger point of the running top Yukawa
 and the SU(2) gauge couplings.
Three solid lines (in red, black and blue) from top to bottom
 correspond to top quark pole masses $M_t=$172.7, 170.9 and 169.1 GeV,
 respectively.
}
%
\end{figure}

\begin{figure}[t,width=12cm, height=8cm]
\begin{center}
{\includegraphics[height=7cm]{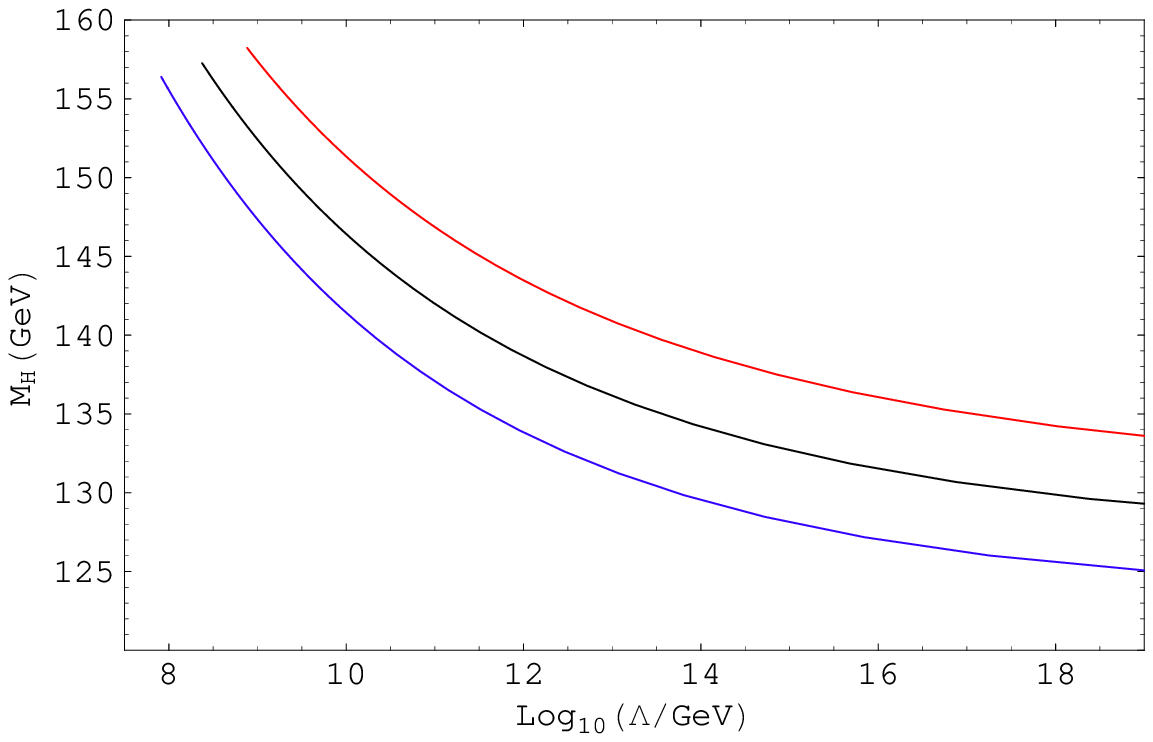} }
\end{center}
\caption{ \small
Plot of Higgs boson mass versus compactification scale,
 $\mbox{Log}_{10}( \Lambda /\mbox{GeV})$,
 determined as the unification scale
 of the running top Yukawa and the SU(2) gauge couplings.
Three solid lines (in red, black and blue) from top to bottom
 correspond to top pole masses $M_t=$172.7, 170.9 and 169.1 GeV,
 respectively.
}%
\end{figure}

\begin{figure}[t, width=12cm, height=8cm]
\begin{center}
{\includegraphics[height=7cm]{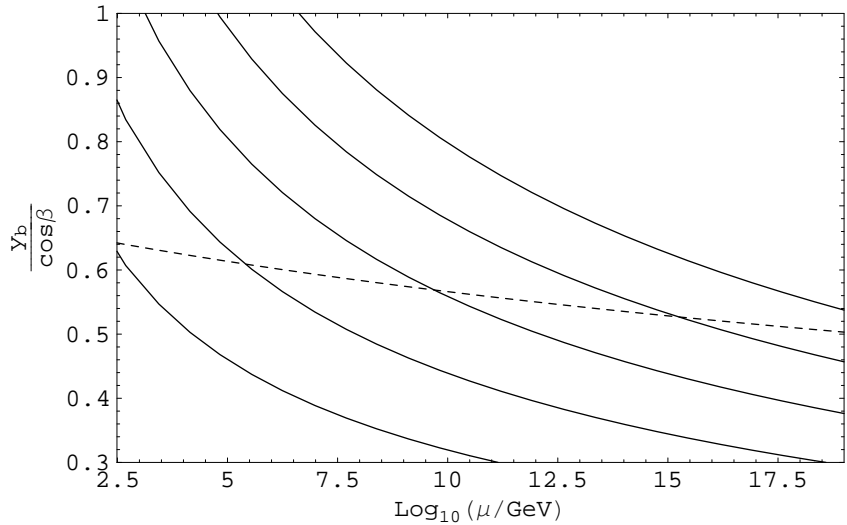} }
\end{center}
\caption{ \small
Plot of the running SU(2) gauge coupling $g(\mu)$ (dashed line)
 and the running bottom Yukawa coupling
 (divided by a given $\cos\beta$), $y_b(\mu)/\cos \beta$ (solid line),
 versus $\mbox{Log}_{10}( \mu /\mbox{GeV})$ for a given $\tan\beta$.
Solid lines correspond to $\tan \beta=40$, 55, 70, 85, and 100, respectively,
 from bottom to top. Here we took the top quark pole mass $M_t=170.9$ GeV.
}
%
\end{figure}

\begin{figure}[t, width=12cm, height=8cm]
\begin{center}
{\includegraphics[height=7cm]{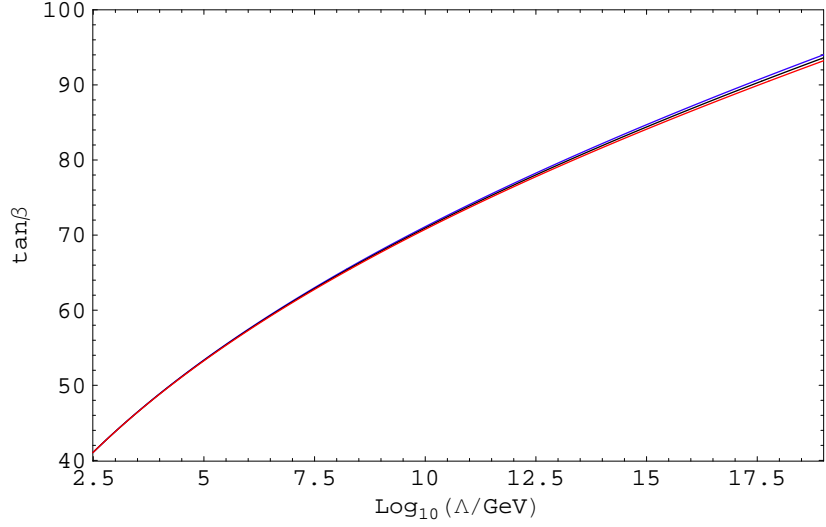} }
\end{center}
\caption{ \small
Plot of $\tan \beta$ versus the compactification scale,
 $\mbox{Log}_{10}( \Lambda /\mbox{GeV})$,
 determined as the unification scale of the running bottom Yukawa
 and the SU(2) gauge couplings.
Three solid line (in blue, black and red) from top to bottom
 correspond to top quark pole masses $M_t=$169.1, 170.9 and 172.7 GeV,
 respectively.
Three lines are well degenerate because their differences originate
 from the running of gauge coupling from $M_Z$ to $M_t$.
}
%
\end{figure}

\begin{figure}[t,width=12cm, height=8cm]
\begin{center}
{\includegraphics[height=7cm]{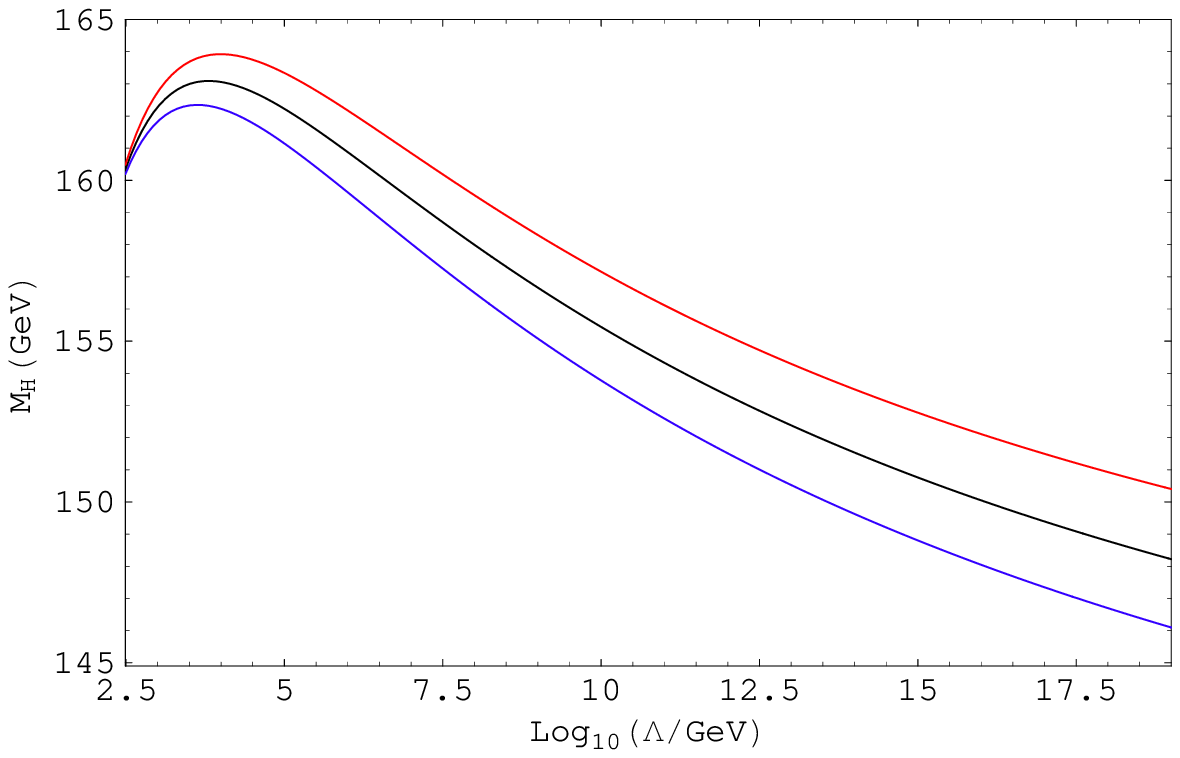} }
\end{center}
\caption{ \small
Plot of Higgs boson mass versus compactification scale,
 $\mbox{Log}_{10}( \Lambda/\mbox{GeV})$,
 determined as the unification scale
 of the running bottom Yukawa and the SU(2) gauge couplings.
Three solid lines (in red, black and blue) from top to bottom
 correspond to top pole masses $M_t=$172.7, 170.9 and 169.1 GeV,
 respectively.
}%
\end{figure}


\begin{thebibliography}{99}


\bibitem{GOS}
I.~Gogoladze, N.~Okada and Q.~Shafi,
  arXiv:0705.3035 [hep-ph],
  to be published in Phys. Lett. B.


\bibitem{GHUcond}
N.~Haba, S.~Matsumoto, N.~Okada and T.~Yamashita,
 JHEP {\bf 0602}, 073 (2006).


\bibitem{Tevatron}
    [CDF Collaboration],
  arXiv:hep-ex/0703034.


\bibitem{6DGHU1}
I.~Antoniadis, K.~Benakli and M.~Quiros,
 New J.\ Phys.\  {\bf 3}, (2001), 20.


\bibitem{6DGHU2}
C.~A.~Scrucca, M.~Serone, L.~Silvestrini and A.~Wulzer,
  JHEP {\bf 0402}, 049 (2004).


\bibitem{RGE}
M.~E.~Machacek and M.~T.~Vaughn,
Nucl.\ Phys.\ B {\bf 222}, 83 (1983);
Nucl.\ Phys.\ B {\bf 236}, 221 (1984);
Nucl.\ Phys.\ B {\bf 249}, 70 (1985);
C.~Ford, I.~Jack and D.~R.~T.~Jones,
  Nucl.\ Phys.\  {\bf B387} (1992) 373,
  [Erratum-ibid.\  {\bf B504} (1997)  551];
V.~D.~Barger, M.~S.~Berger and P.~Ohmann,
  Phys.\ Rev.\  D {\bf 47}, 1093 (1993);
  M.~X.~Luo and Y.~Xiao,
  Phys.\ Rev.\ Lett.\  {\bf 90}, 011601 (2003).


\bibitem{PDG}
W.~M.~Yao {\it et al.}  [Particle Data Group],
 J.\ Phys.\ G {\bf 33} (2006) 1.


\bibitem{Pole-MSbar}
See, for example, H.~Arason, D.~J.~Castano, B.~Keszthelyi,
S.~Mikaelian,
 E.~J.~Piard, P.~Ramond and B.~D.~Wright,
     Phys.\ Rev.\  D {\bf 46}, 3945 (1992);
%
H.~E.~Haber, R.~Hempfling and A.~H.~Hoang,
  Z.\ Phys.\  C {\bf 75}, 539 (1997).


\bibitem{Pole-MSbar2}
See, for example,
F.~Jegerlehner, M.~Y.~Kalmykov and O.~Veretin,
  Nucl.\ Phys.\  B {\bf 641}, 285 (2002);
  Nucl.\ Phys.\  B {\bf 658}, 49 (2003);
%
F.~Jegerlehner and M.~Y.~Kalmykov,
  Nucl.\ Phys.\  B {\bf 676}, 365 (2004).


\bibitem{LEP2}
R.~Barate {\it et al.}  [LEP Working Group for Higgs boson
searches],
 Phys.\ Lett.\  B {\bf 565}, 61 (2003).


\bibitem{stability1}
N.~Cabibbo, L.~Maiani, G.~Parisi and R.~Petronzio,
  Nucl.\ Phys.\  B {\bf 158}, 295 (1979);
  P.~Q.~Hung,
  Phys.\ Rev.\ Lett.\  {\bf 42}, 873 (1979);
  M.A.B. Beg, C. Panagiotakopoulos and A. Sirlin, Phys. Rev. Lett.52 (1984)
  883;
M.~Lindner,
  Z.\ Phys.\  C {\bf 31}, 295 (1986);
 M.~Sher,
  Phys.\ Rept.\  {\bf 179}, 273 (1989);
G.~Altarelli and G.~Isidori,
 Phys.\ Lett.\  B {\bf 337}, 141 (1994);
%
J.~A.~Casas, J.~R.~Espinosa and M.~Quiros,
 Phys.\ Lett.\  B {\bf 342}, 171 (1995);
%
 Phys.\ Lett.\  B {\bf 382}, 374 (1996);
%
J.~R.~Espinosa and M.~Quiros,
  Phys.\ Lett.\  B {\bf 353}, 257 (1995).


\bibitem{Csaki}
For example, see G.~Cacciapaglia, C.~Csaki and S.~C.~Park,
 JHEP {\bf 0603}, 099 (2006),
and  references therein.


\bibitem{GHY}
I.~Gogoladze, Y.~Mimura and S.~Nandi,
Phys.\ Lett.\  B {\bf 560}, 204 (2003);
Phys.\ Lett.\ B {\bf 562}, 307 (2003);
Phys.\ Rev.\ Lett.\  {\bf 91}, 141801 (2003);
Phys.\ Rev.\ D {\bf 69}, 075006 (2004);
I.~Gogoladze, T.~Li, Y.~Mimura and S.~Nandi,
  Phys.\ Rev.\  D {\bf 72}, 055006 (2005);
I.~Gogoladze, C.~A.~Lee, Y.~Mimura and Q.~Shafi,
  arXiv:hep-ph/0703107.


\bibitem{FuKo}
See, for example,
H.~Fusaoka and Y.~Koide,
 Phys.\ Rev.\  D {\bf 57}, 3986 (1998).


\bibitem{NCY1}
  V.~Barger, J.~Jiang, P.~Langacker and T.~Li,
  Phys.\ Lett.\  B {\bf 624}, 233 (2005);
  Nucl.\ Phys.\  B {\bf 726}, 149 (2005);
  Model with non-canonical normalization for $U(1)_Y$
  have also been discussed by I.~Gogoladze, T.~Li, V.~N.~Senoguz and Q.~Shafi,
  Phys.\ Rev.\  D {\bf 74}, 126006 (2006); I.~Gogoladze, C.~A.~Lee, T.~Li and Q.~Shafi,
  arXiv:0704.3568 [hep-ph];  A. Kehagias and N.D. Tracas,
  arXiv:hep-ph/0506144; A.~Aranda, J.~L.~Diaz-Cruz and A.~Rosado,
 Int.\ J.\ Mod.\ Phys.\  A {\bf 22}, 1417 (2007).


\bibitem{NCU}
I.~Gogoladze, T.~Li and Q.~Shafi,
 Phys.\ Rev.\  D {\bf 73}, 066008 (2006)


\bibitem{Antoniadis:1990ew}
  I.~Antoniadis,
  Phys.\ Lett.\  B {\bf 246}, 377 (1990).


\bibitem{UED}
T.~Appelquist, H.~C.~Cheng and B.~A.~Dobrescu,
  Phys.\ Rev.\  D {\bf 64}, 035002 (2001).


\bibitem{UED-DM}
H.~C.~Cheng, K.~T.~Matchev and M.~Schmaltz,
 Phys.\ Rev.\  D {\bf 66}, 036005 (2002).


\bibitem{LKP}
G.~Servant and T.~M.~P.~Tait,
 Nucl.\ Phys.\  B {\bf 650}, 391 (2003);
%
M.~Kakizaki, S.~Matsumoto, Y.~Sato and M.~Senami,
 Phys.\ Rev.\  D {\bf 71}, 123522 (2005);
 Nucl.\ Phys.\  B {\bf 735}, 84 (2006);
%
K.~Kong and K.~T.~Matchev,
 JHEP {\bf 0601}, 038 (2006);
%
F.~Burnell and G.~D.~Kribs,
 Phys.\ Rev.\  D {\bf 73}, 015001 (2006);
%
M.~Kakizaki, S.~Matsumoto and M.~Senami,
 Phys.\ Rev.\  D {\bf 74}, 023504 (2006);
See, also,
B.~A.~Dobrescu, D.~Hooper, K.~Kong and R.~Mahbubani,
 arXiv:0706.3409 [hep-ph],
where the scalar component of photon
 in a six-dimensional UED model
 is the dark matter candidate.

\end{thebibliography}
\end{document}